\documentclass[preprint,12pt]{elsarticle}

\usepackage[dvips]{graphicx}
\usepackage{epstopdf}
\usepackage{amssymb}
\usepackage{amsmath}
\usepackage{subfigure}
\usepackage{color}
\usepackage{multirow}
\journal{Physica A}

\begin{document}

\begin{frontmatter}

\title{Diffusion-like recommendation with enhanced similarity of objects}


\author[inst1]{Ya-Hui An}
\author[inst1,inst2]{Qiang Dong\corref{cor1}}
\author[inst2]{Chong-Jing Sun}
\author[inst1]{Da-Cheng Nie}
\author[inst1,inst2]{Yan Fu}
\cortext[cor1]{Corresponding author.} \ead{dongq@uestc.edu.cn}
\address[inst1]{School of Computer Science $\&$ Engineering, University of Electronic Science and Technology of China, Chengdu 611731, P.R. China}
\address[inst2]{Big Data Research Center, University of Electronic Science and Technology of China, Chengdu 611731, P.R. China}

\begin{abstract}
In the last decade, diversity and accuracy have been
regarded as two important measures in evaluating a recommendation
model. However, a clear concern is that a model focusing excessively
on one measure will put the other one at risk, thus it is not easy
to greatly improve diversity and accuracy simultaneously. In this
paper, we propose to enhance the Resource-Allocation (RA) similarity
in resource transfer equations of diffusion-like models, by giving a
tunable exponent to the RA similarity, and traversing the value of
this exponent to achieve the optimal recommendation results. In this
way, we can increase the recommendation scores (allocated resource)
of many unpopular objects. Experiments on three benchmark data sets,
MovieLens, Netflix and RateYourMusic show that the modified models
can yield remarkable performance improvement compared with the
original ones.
\end{abstract}


\end{frontmatter}


\section{Introduction}
\label{Introduction}
Nowadays, the explosive growth of storage capability has allowed
people to collect and store almost all the information generated
every day. This so-called \emph{"big data"} provides great benefit
to our life, e.g., we can easily get the cast list of a niche film
via search engines. However, we find that it becomes very
difficult to find the relevant movie of our interest from countless
candidates, if we cannot describe it by appropriate keywords. On
this \emph{information overload} occasion, \emph{recommender
systems} arise to help us make the right decisions.

Different from search engines requiring keywords, recommender
systems are designed to uncover users' potential preferences and
interests based on users' past activities and profile descriptions,
and accordingly deliver a personalized list of recommended objects
to every user. In the last decade, recommender systems have become a
significant issue in both academic and industrial communities. The
early recommender models are based on a simple observation that
similar users are likely to purchase the same items, or, the items
collected by the same user are prone to be similar to each other,
such as collaborate filtering \cite{26} and content-based methods \cite{27}. These
methods are shown to give accurate recommendation results, but they confront a recommender system with the risk that more and more
users will be exposed to a narrowing band of popular items, leading
to poor diversity among users' recommendation lists \cite{9,12}.

Given this, many other recommendation models have been proposed in
the literature, including dimensionality reduction techniques
\cite{1,13,14,15}, diffusion-like methods \cite{2,16,19,25}, social
filtering \cite{17,18,23}, and hybrid recommendation models
\cite{1,11,22,24}. However, people found that accuracy and diversity
seem to be two sides of the seesaw: when one side rises, the other
side falls. Examples are two primary diffusion-like methods, ProbS
\cite{2} and HeatS \cite{3}, which mimic two basic physical
processes on user-object bipartite networks. ProbS is demonstrated
to give recommendation results with good accuracy but poor
diversity, while HeatS is found to be effective in providing a
diverse recommendation lists at the cost of accuracy.

This diversity-accuracy dilemma has received considerable research
attentions in the field of recommender systems. Zhou et al. \cite{1}
designed delicately a nonlinear hybrid model of HeatS and ProbS,
called HHP, which achieves significant improvements in both accuracy
and diversity of recommendation results. Another two effective
methods modified respectively from original ProbS and HeatS, named
Preferential Diffusion (PD) \cite{5} and Biased Heat Conduction
(BHC) \cite{4}, also make a good trade off on accuracy and
diversity. In addition, based on BHC and HHP, Qiu et al. \cite{21} took the heterogeneity of the source objects into account, and proposed the Heterogeneous Heat Conduction(HHC) model and Nie et al. \cite{19} investigated the
optimal hybrid coefficients of HeatS and ProbS, and proposed
accordingly the Balance Diffusion (BD) model, respectively.

In the first diffusion-like model ProbS, every user distributes the
total resource he receives previously from objects, back averagely
to his neighbor objects. The niche objects will receive lower final
resources (recommendation scores) because they have fewer neighbor
users (resource portals), thus rank in the bottom of the
recommendation lists. That is why ProbS suffers from poor diversity.
In view of this, the PD model proposed by Lv et al. \cite{5}
intentionally allocates more resource to small-degree objects, and
less resource to large-degree objects. For the resource of a given
user, every neighbor object receives the percentage approximately
inversely proportional to its degree. Compared with ProbS, PD
simultaneously improves the diversity and accuracy of recommendation
results.

With the similar motivation, we proposed a method which can be used
in many being diffusion-like algorithms, like ProbS, BHC and HHP, to
get better recommendation results on user-object bipartite networks.
we propose to enhance the RA similarity in the transfer equations of
diffusion-like models, by giving a tunable exponent on the shoulder
of RA similarity, and traverse this parameter to achieve the optimal
recommendation results. Experiments on three benchmark data sets,
MovieLens, Netflix, and RYM (Rate Your Music) show that our model
can yield a great performance improvement compared with the original
models.

\section{Materials and Methods}
In this paper, a recommender system is represented by a bipartite
network $G(U,O,E)$, where $U = \{u_1,u_2,\cdots,u_m\}$, $O =
\{o_1,o_2, \cdots ,o_n\}$ and $E = \{e_1,e_2, \cdots ,e_q \}$
correspond to $m$ users, $n$ objects and $q$ edges between users and
objects, respectively. This bipartite network could be fully
described by an adjacency matrix $A = \{a_{l\alpha}\}_{m\times n}$,
where the element $a_{l\alpha} = 1$ if there exists an edge between
user $u_l$ and object $o_\alpha$ (user $u_l$ \emph{collects} object
$o_\alpha$), meaning that user $u_l$ declared explicitly his
preference on object $o_\alpha$ in the past, and $a_{l\alpha} = 0$
otherwise. For every target user, the essential task of a
recommender system becomes to recommend him a sublist of uncollected
objects of his potential interest.

\subsection{Dataset Description}
Three real-world data sets are adopted to test the recommendation
result, namely, MovieLens, Netflix and RYM (Rate Your Music). Here
we will briefly describe these three data sets. MovieLens, a movie
rating data set, was collected by the GroupLens Research Project at
the University of Minnesota and can be found at the website
www.grouplens.org. Netflix, a randomly sampled subset of the huge
data set provided by the Netflix company for the Netflix Prize
(www.netflixprize.com) \cite{6}. RYM, a music rating data set, is
obtained by downloading publicly available data from the music
ratings website www.RateYourMusic.com \cite{1}. In this paper, we
make use of nothing but the binary information whether there exists
an interaction between a user and an object in the past. The basic
statistics of the three data sets are presented in
Table.\ref{tab:1}.

\begin{table}
\caption{The basic statistics of three data sets, where $n$, $m$ and
$q$ denote the number of users, objects and edges, respectively;
$\langle{k_u}\rangle$ and $\langle{k_o}\rangle$ are the average
degrees of users and objects.} \label{tab:1}
\begin{center}
\begin{tabular}[b]{l c c c c r}
\hline
Data set& $n$ & $m$ & $q$ & $\langle{k_u}\rangle$ & $\langle{k_o}\rangle$ \\
\hline
MovieLens & 943 & 1,682 & 100,000 & 106 & 59.5 \\
Netflix & 10,000 & 5,640 & 701,947 & 70.2 & 124.5 \\
RYM & 33,762 & 5,267 & 675,817 & 20 & 128.3 \\
\hline
\end{tabular}
\end{center}
\end{table}
To evaluate the performance of different models, each data set is
randomly divided into two subsets: the training set $E^T$ containing
$90\%$ of the links and the probe set $E^P$ with $10\%$ of the
links. The training set is treated as known information to make
recommendation and the probe set is only used to test the relevance
of the recommendation results.

\subsection{Evaluation Metrics}
In order to evaluate the relevance of recommendation results, we
adopt three typical metrics in this paper.  In recommender systems,
$accuracy$ is the most important aspect in evaluating the
recommendation performance. A good algorithm is expected to give
accurate recommendations, namely stronger ability to find what the
users like. We make use of ranking score $RS$ \cite{4} and precision
enhancement $ep(L)$ \cite{1} to measure the recommendation accuracy.
For a target user, the recommender system will return a ranking list
of all his uncollected objects to him. For a link between user $u_i$
and object $o_\alpha$ in probe set, we compute the rank ($RS_{i
\alpha}$) of object $o_\alpha$ in the recommendation list of user
$u_i$.
\begin{center}
\begin{equation}
\label{equ4} RS_{i\alpha} = \frac{p_{\alpha}}{l_i},
\end{equation}
\end{center}

In the above definition, object $\alpha$ is ranked in the
$p_{\alpha}$-th position of the recommendation list of user $i$, and
$ l_i = n - k^{T}_{i}$ is the number of objects in recommendation
list of user $i$, where $k^{T}_{i}$ is the degree of user $i$ in the
training set $E^T$. The ranking score $RS$ of the whole system can
be obtained by averaging the ranking score values over all
user-object links in the probe set.

In practice, we will focus on the top-ranked objects rather than
checking the whole recommendation list. Thus, a more practical
approach is to check the top $L$ objects recommended to the target
user to calculate how many objects are recommended correctly.
$Precision$ is the mostly used metric defined in this way. However,
for a sparse data sets the precision may be very low, while for a
dense data set it may be high. Obviously, finding a better way to
avoid some bad effects brought from data sets themselves is
necessary. Zhou et al. \cite{1} introduced enhanced precision
$ep(L)$ which considers improvement compared with the precision of
random recommendations. A random recommendation will randomly choose
$L$ objects from the train set and recommend them to the target
user, where $L$ is the length of the recommendation list.

\begin{center}
\begin{equation}
\label{equ6} ep(L) =
\frac{1}{m^P}\sum_{l=1}^{m^P}\frac{n_l}{L}\times\frac{n^T-k^T_l}{k^P_l},
\end{equation}
\end{center}

where $m^P$ is the number of users in the probe set, $L$ is the
length of recommendation list, $n_l$ is the number of relevant
objects in the recommendation list of user $u_l$, $n^T$ is the
number of objects in the train set, $n^T-k^T_l$ is the number of
uncollected objects of user $l$ in the train sets, and $k^P_l$ is
the degree of user $u_l$ in probe set.

Besides accuracy, the personalized recommendation method should
present different recommendations to different users according to
their tastes and interests. The diversity can be quantified by the
average Hamming Distance, which measures how different are the
recommended lists of users from each other. Given two different
users $u_i$ and $u_j$, borrowing inspiration from the Hamming
distance between two strings, the diversity is calculate in a
similar way \cite{20},

\begin{center}
\begin{equation}
\label{equ7} h_{ij}(L) = 1 - \frac{q_{ij}(L)}{L},
\end{equation}
\end{center}

where $q_{ij}$ is the number of common objects in the top $L$
positions of both lists of user $u_i$ and user $u_j$. Clearly, if
user $i$ and user $j$ receive the same recommendation list,
$h_{ij}(L) = 0$, while if their lists are completely different,
$h_{ij}(L) = 1$. Averaging $h_{ij}(L)$ over all pairs of active
users in the probe set, we obtain the hamming distance $h(L)$ of the
whole system, where greater value means better personalization of
users' recommendation lists.

\subsection{Diffusion-like recommendation models}
Most diffusion-like recommendation models work by assigning every
object $o_{\alpha}$ an initial level of resource $f_{\alpha}$, where
the resources on all the objects constitute a resource vector
$\textbf{f}$; then the resources will be redistributed among objects
according to the formula $\textbf{f}' = W\textbf{f}$ .

Analogous to mass diffuse process in user-object bipartite network,
Zhou et al. proposed the Probabilistic Spreading (ProbS) model
\cite{2}, also referred to as Network-Based Inference (NBI). For a
target user $u_l$, the initial resource vector $\textbf{f}$ is
defined as $f_{\alpha} = a_{l\alpha}$, where $a_{l\alpha}=1$ if user
$u_l$ has collected object $o_\alpha$, otherwise $a_{l\alpha}=0$.
The transfer equation $w_{\alpha\beta}$ in matrix $W$ is written as

\begin{center}
\begin{equation}
\label{equ1} w_{\alpha \beta}^{\text{ProbS}} =
\frac{1}{k_{o_{\beta}}}
\sum_{l=1}^{m}{\frac{a_{l\alpha}a_{l\beta}}{k_{u_l}}},
\end{equation}
\end{center}
where $k_{o_{\beta}} = \sum_{i = 1}^{m}{a_{i\beta}}$ and $k_{u_l} =
\sum_{r = 1}^{n}{a_{lr}}$ denote the degrees of object $o_{\beta}$
and user $u_l$, respectively.

Another diffusion-like model mimicking the heat-spreading process is
called HeatS \cite{3}. The initial resource vector $\textbf{f}$ of
HeatS is the same as that of ProbS. The key difference between ProbS
and HeatS is the resource redistribution strategy: ProbS works by
equally distributing the resource of each node to all of its nearest
neighbors, the overall resource remains unchanged; while in HeatS
every node absorbs equal proportion of resource from every nearest
neighbor, the overall resource increases in the process.
Specifically, the difference of HeatS from ProbS lies in the
transfer matrix $W$, which is described as:

\begin{center}
\begin{equation}
\label{equ2} w_{\alpha \beta}^{\text{HeatS}} =
\frac{1}{k_{o_{\alpha}}}
\sum_{l=1}^{m}{\frac{a_{l\alpha}a_{l\beta}}{k_{u_l}}},
\end{equation}
\end{center}

As we mentioned before, ProbS enjoys high recommendation accuracy
yet low diversity, while HeatS designed specifically to address the
challenge of diversity suffers from terrible accuracy. Many
researchers attempted to solve this diversity-accuracy dilemma and
have found out some effective ways. For example, a \emph{hybrid}
model of HeatS and ProbS, named HHP, was proposed, with a tunable
parameter $\lambda$ in the transfer equation $w_{\alpha \beta}$:

\begin{center}
\begin{equation}
\label{equ3} w_{\alpha \beta}^{\text{HHP}} =
\frac{1}{{k_{o_{\alpha}}}^{1-\lambda} {k_{o_{\beta}}}^{\lambda}}
\sum_{l=1}^{m}{\frac{a_{l\alpha }a_{l\beta}}{k_{u_l}}},
\end{equation}
\end{center}

It is easily observed that HHP reduces to ProbS when $\lambda = 1$,
and HeatS when $\lambda = 0$.

\section{Results}
\subsection{Empirical analysis}

\begin{figure*}
\centering \subfigure[MovieLens]{
\begin{minipage}[b]{0.3\textwidth}
\includegraphics[width=1\textwidth]{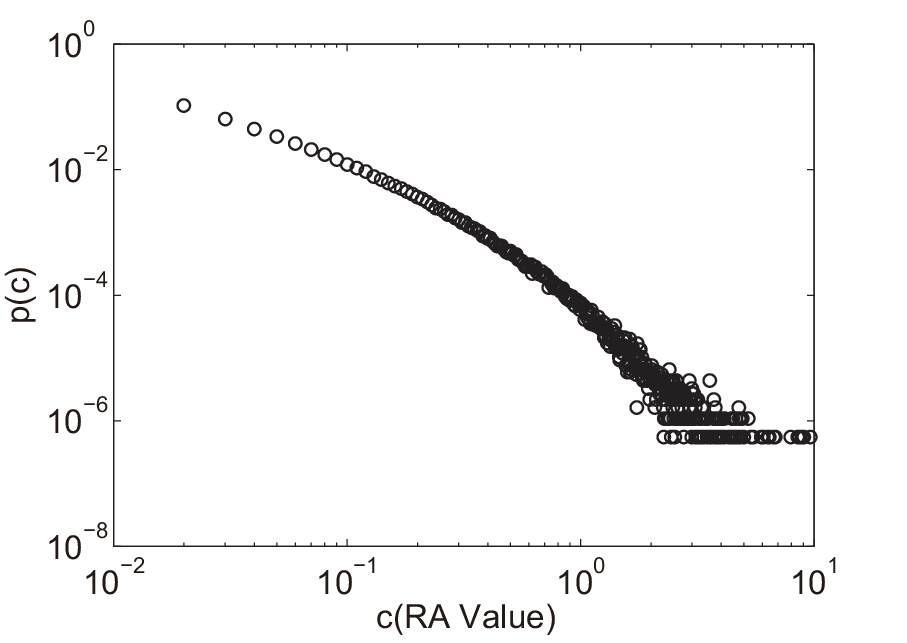}
\end{minipage}
} \subfigure[Netflix]{
\begin{minipage}[b]{0.3\textwidth}
\includegraphics[width=1\textwidth]{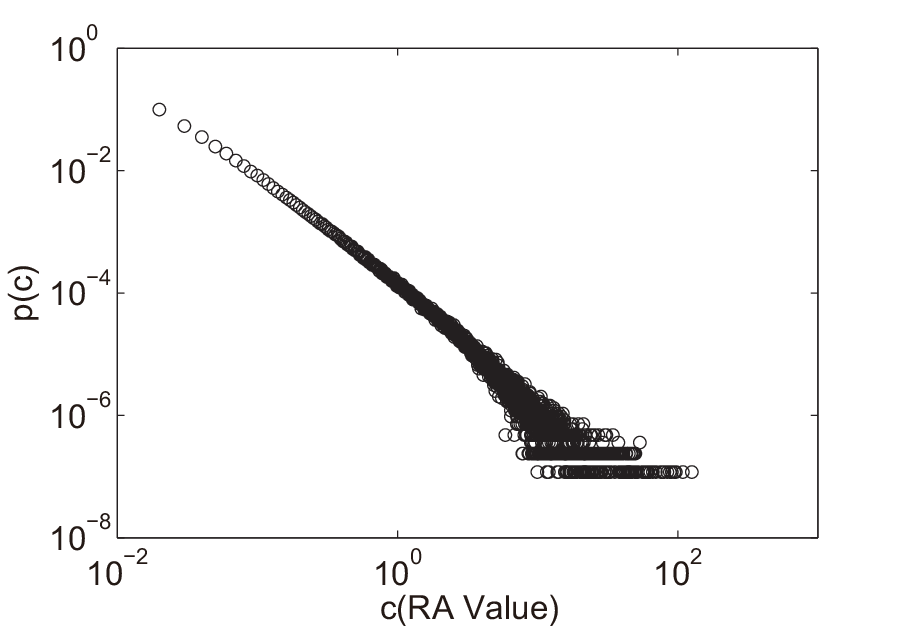}
\end{minipage}
} \subfigure[RYM]{
\begin{minipage}[b]{0.3\textwidth}
\includegraphics[width=1\textwidth]{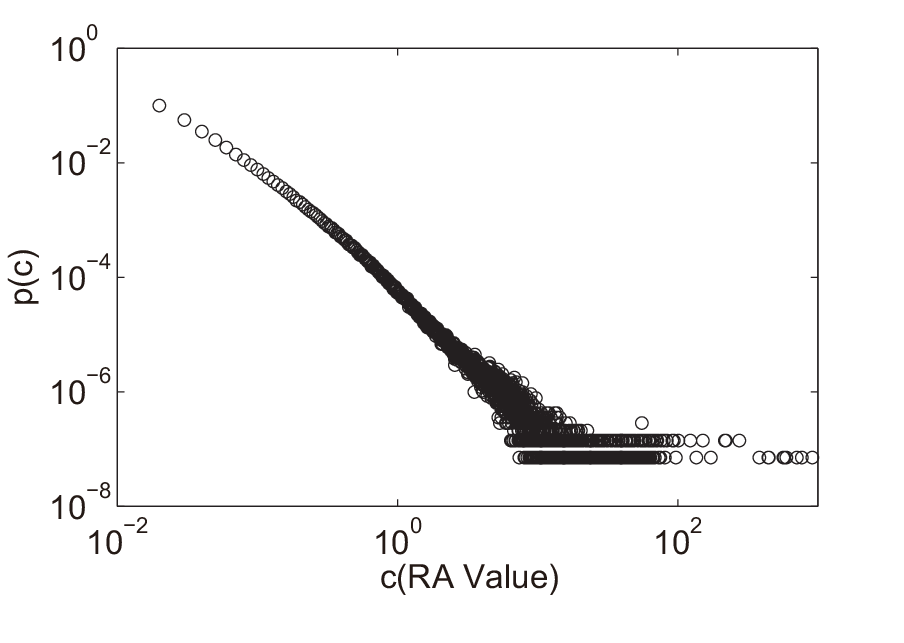}
\end{minipage}
} \caption{The power-law distribution of RA values on three
different data sets. }\label{fig:1}
\end{figure*}

For most of aforementioned diffusion-like methods, the summation
formula $\sum_{l=1}^{m}{\frac{a_{l\alpha }a_{l\beta}}{k_{u_l}}}$ is
a common component of the transfer equations. Zhou et al. \cite{10}
defined this summation formula as the \emph{Resource-allocation
(RA)} index, which is regarded as a significant similarity measure
of two objects because it considers not only the number of common
users but also the degrees of these common users. Fig.\ref{fig:1}
shows the distributions of RA index on three data sets. It is clear
that the distributions follow almost power-law, which indicates that
the RA similarities of most pairs of objects are weak and the RA
value ranges a large scope (see Table.\ref{tab:ss}).

\begin{figure*}
\centering \subfigure[RA on MovieLens]{
\begin{minipage}[b]{0.3\textwidth}
\includegraphics[width=1\textwidth]{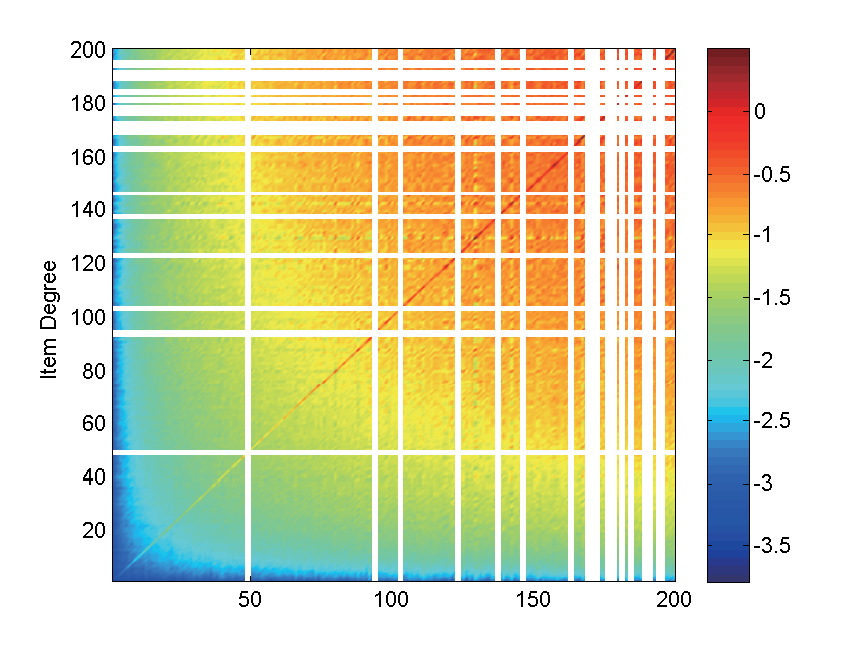}
\end{minipage}
} \subfigure[RA on Netflix]{
\begin{minipage}[b]{0.3\textwidth}
\includegraphics[width=1\textwidth]{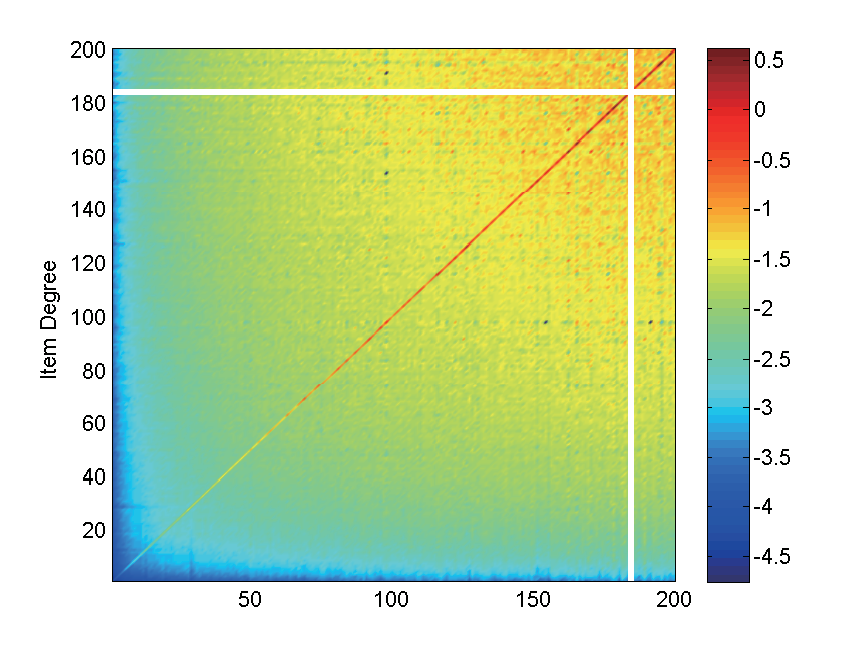}
\end{minipage}
} \subfigure[RA on RYM]{
\begin{minipage}[b]{0.3\textwidth}
\includegraphics[width=1\textwidth]{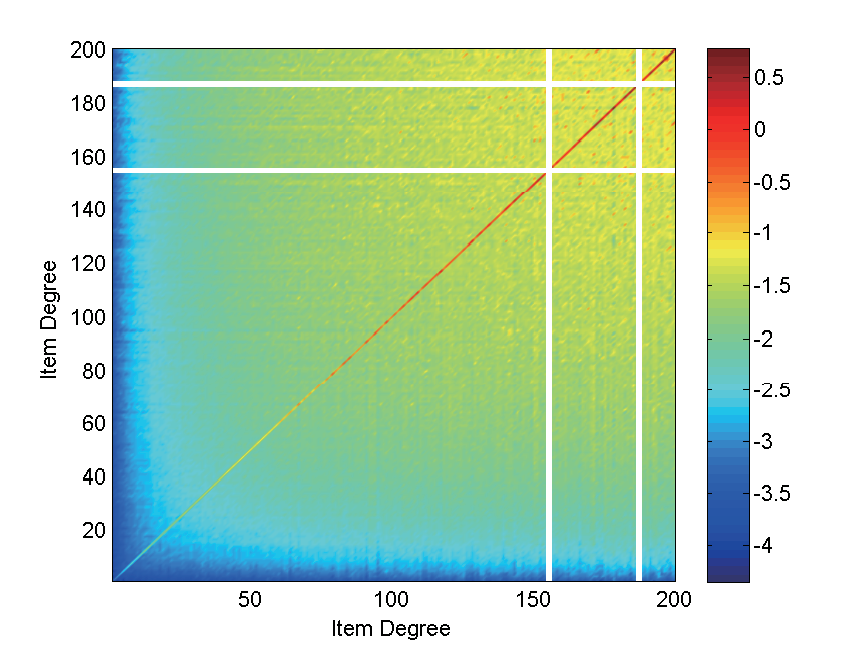}
\end{minipage}
}

\subfigure[ERA on MovieLens]{
\begin{minipage}[b]{0.3\textwidth}
\includegraphics[width=1\textwidth]{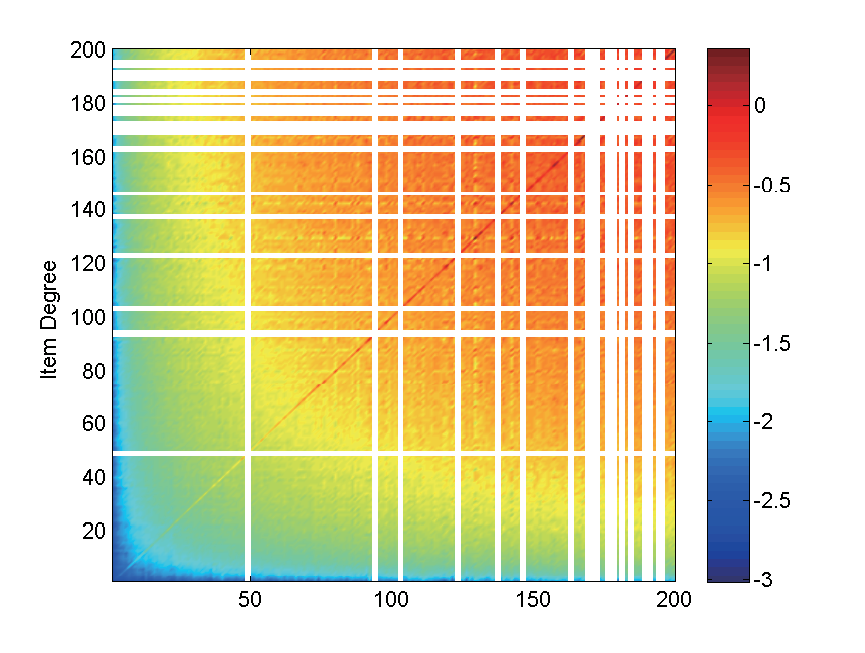}
\end{minipage}
} \subfigure[ERA on Netflix]{
\begin{minipage}[b]{0.3\textwidth}
\includegraphics[width=1\textwidth]{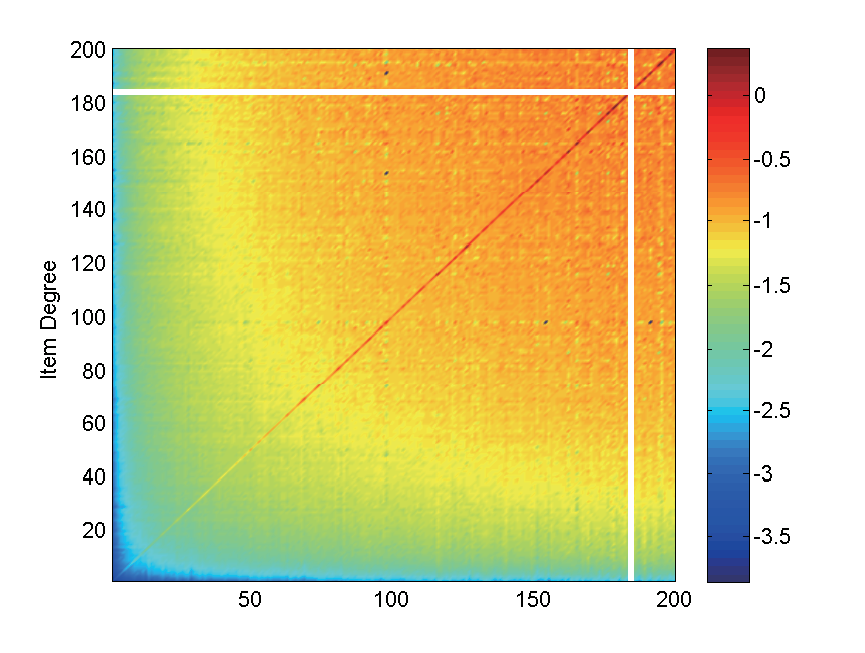}
\end{minipage}
} \subfigure[ERA on RYM]{
\begin{minipage}[b]{0.3\textwidth}
\includegraphics[width=1\textwidth]{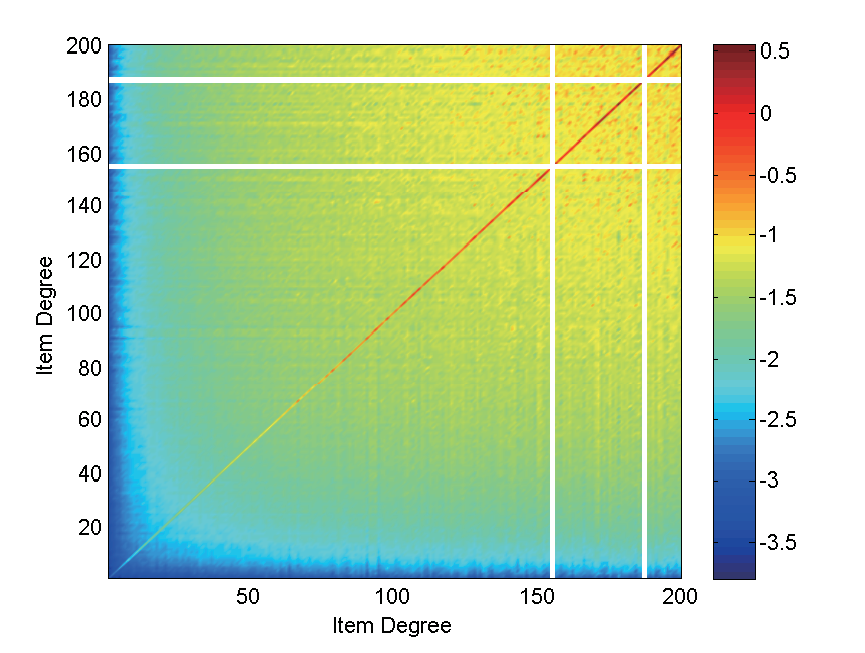}
\end{minipage}
}

\caption{The heat map of RA and ERA similarities against degrees of
arbitrary pair of objects on three data sets. }\label{fig:2}
\end{figure*}

For the user-object bipartite network, the RA similarity can not
effectively distinguishes the similarity of objects in an
appropriate scale. Fig.\ref{fig:2} plots the heat map of RA
similarity against degrees of arbitrary pair of objects on three
data sets (Fig.2(a), Fig.2(b), Fig.2(c)). the darker is the color,
the bigger is the RA similarity of two objects. We find that a
majority of RA values are very low, especially for the pairs of
small-degree objects, which validates again the heavy-tailed
distribution of RA values.

\begin{table}
\caption{The minimum and maximum values of RA index on three data
sets.} \label{tab:ss}
\begin{center}
\begin{tabular}[b]{l c c r}
\hline
Data set & minRA & maxRA & magnification \\
\hline
MovieLens & 0.0015 & 9.60 & $6.4\times 10^3$ \\
Netflix & 0.0010 & 126.5862 & $1.2\times 10^5$ \\
RYM & 0.00058 & 912.8531 &  $1.57\times 10^6$\\
\hline
\end{tabular}
\end{center}
\end{table}

Note that the RA index is a key factor in measuring the resource
transfer between objects. Based on above empirical results, we infer
that a few large-degree objects receive much resource (recommending
opportunity), while a majority of small-degree objects get little
resource, thus are seldom recommended. What is more, two popular
objects are likely to be bought by the same users, which does not
mean these two objects are similar to each other. On the contrary,
two niche objects bought by the same users may have high similarity.
Thus, the recommendation performance may be improved if we modify
the RA index to intentionally decrease (increase) the similarities
of large-degree (small-degree) objects. Specifically, the negative
effect of power-law distribution of RA values will be weaken by
giving an exponent $\sigma < 1$ on the shoulder of RA index, named
\emph{Enhanced Resource Allocation (ERA)} index:

\begin{center}
\begin{equation}
\label{equ8} s^{\text{ERA}}_{\alpha \beta} =
{(\sum_{l=1}^{m}{\frac{a_{l\alpha }a_{l\beta}}{k_{u_l}}})}^{\sigma},
\end{equation}
\end{center}

By adjusting the parameter $\sigma$ of ERA similarity to appropriate
values, we get the heat map of ERA similarities against the degrees
of objects on three datasets (Fig.\ref{fig:2}), where $\sigma = 0.7$
for MovieLens(Fig.2(d)), $\sigma = 0.6$ for Netflix(Fig.2(e)) and
$\sigma = 0.8$ for RYM(Fig.2(f)). Compared with Fig.2(a), Fig.2(b)
and Fig.2(c) respectively, we find that the similarities of many
pairs of small degree objects are increased from RA index to ERA
index, reflected by larger area of darker color in the figure.

\begin{figure*}[htbp]
\centering
\subfigure[MovieLens]{
\begin{minipage}[b]{0.45\textwidth}
    \includegraphics[width=1\textwidth]{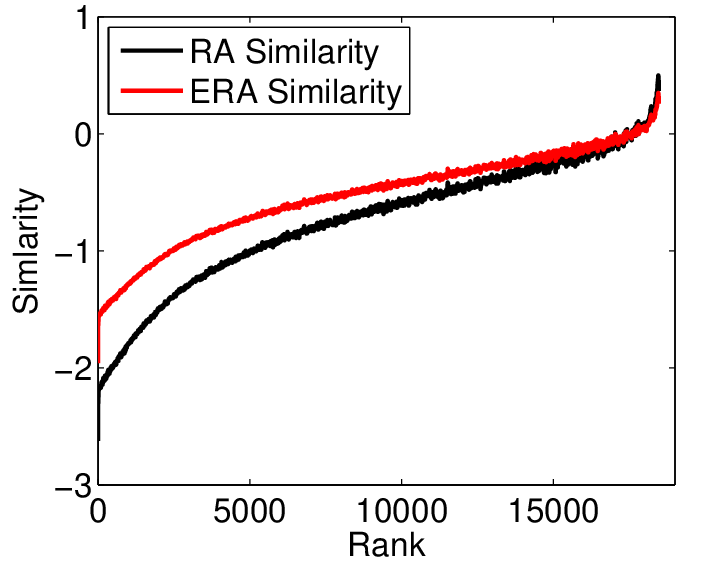}
\end{minipage}
} \subfigure[Netflix]{
\begin{minipage}[b]{0.45\textwidth}
	\includegraphics[width=1\textwidth]{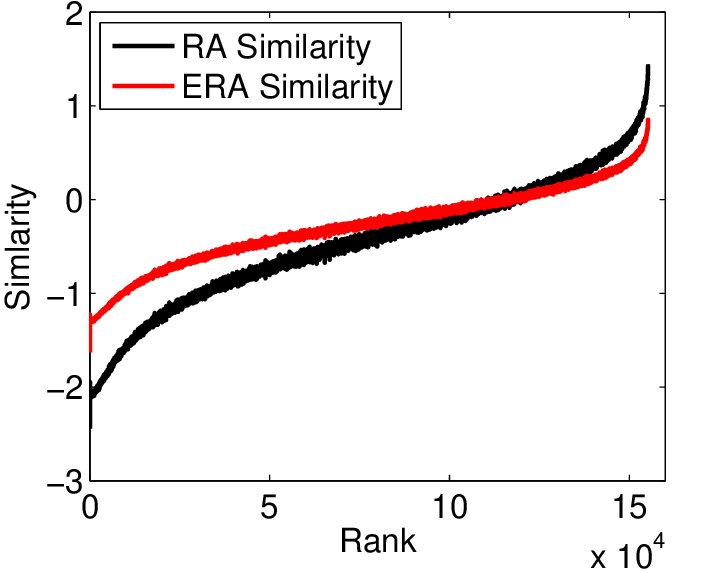}
\end{minipage}
} \subfigure[RYM]{
\begin{minipage}[b]{0.45\textwidth}
	\includegraphics[width=1\textwidth]{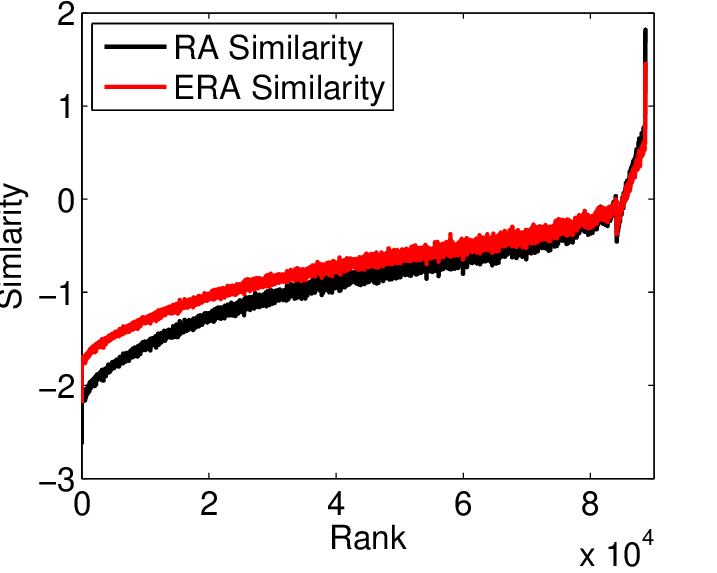}
\end{minipage}
} \caption{The change of RA and ERA similarities with the \textcolor{red}{incresing ranks} of degree products of object pairs. Each data point is obtained by
1-dimensional median filter with the order $N=30$. For a given $x$, its corresponding $Similarity$ is the median of y-values of $[x-N/2,x+N/2+1]$.}\label{fig:222}
\end{figure*}

To quantitatively illustrate the difference of ERA index from RA
index, Fig.3 plots the similarities against the ranks of degree product of
object pairs.For a given x, its corresponding Similarity is the median of y-values of $[x-N/2,x+N/2+1]$, where $N$ is chosen as 30 for a better illustration. We can see a noticeable gap between the similarities
of ERA and RA for the same degree product. Specifically, there
exists an intersection of the two curves for Netflix data set, which
means that ERA similarity is higher (lower) than RA similarity for
the object pairs of small degrees (large degrees). However for
MovieLens and RYM, ERA similarity is higher than RA similarity for
almost all the object pairs regardless of the degree products. Thus,
we conjecture that by replacing RA with ERA similarity, the
improvement of algorithmic performance will be more significant for
Netflix than the other two data sets, which will be validated in
following experiments.

\subsection{Experiment results}

\begin{figure*}[htbp]
\centering
\includegraphics[height=2.5in,width=0.32\textwidth]{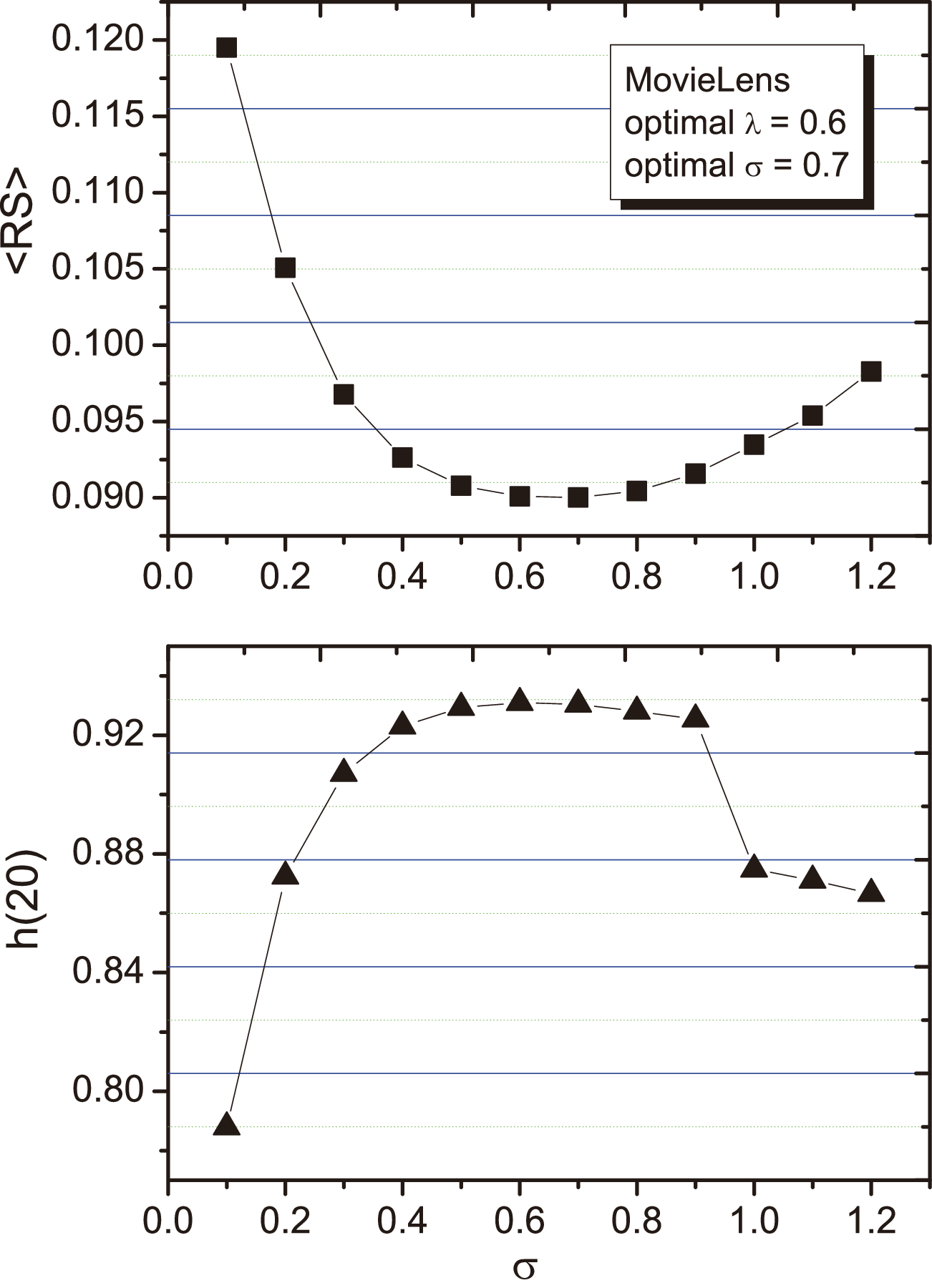}
\hfill
\includegraphics[height=2.5in,width=0.32\textwidth]{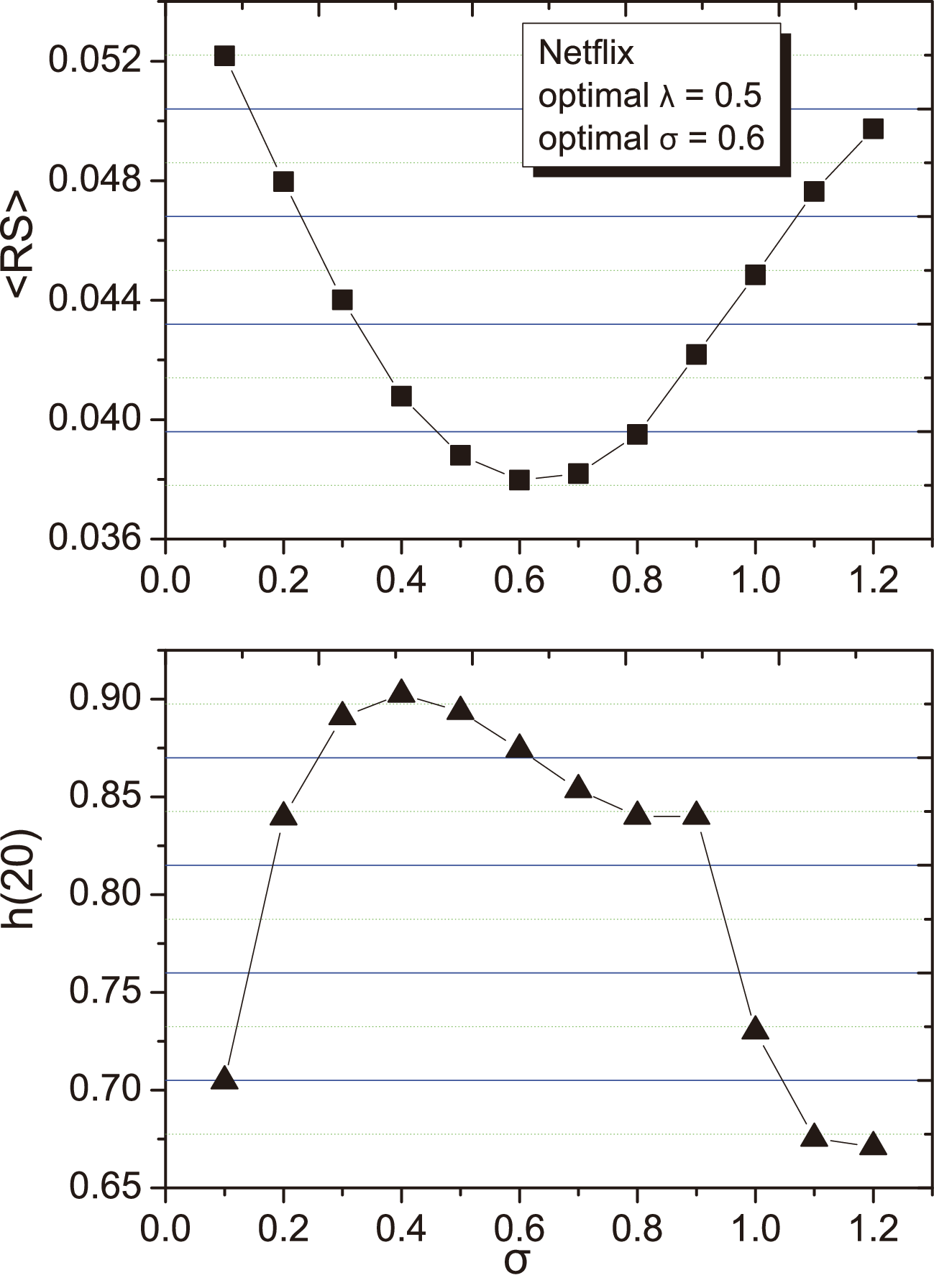}
\hfill
\includegraphics[height=2.5in,width=0.32\textwidth]{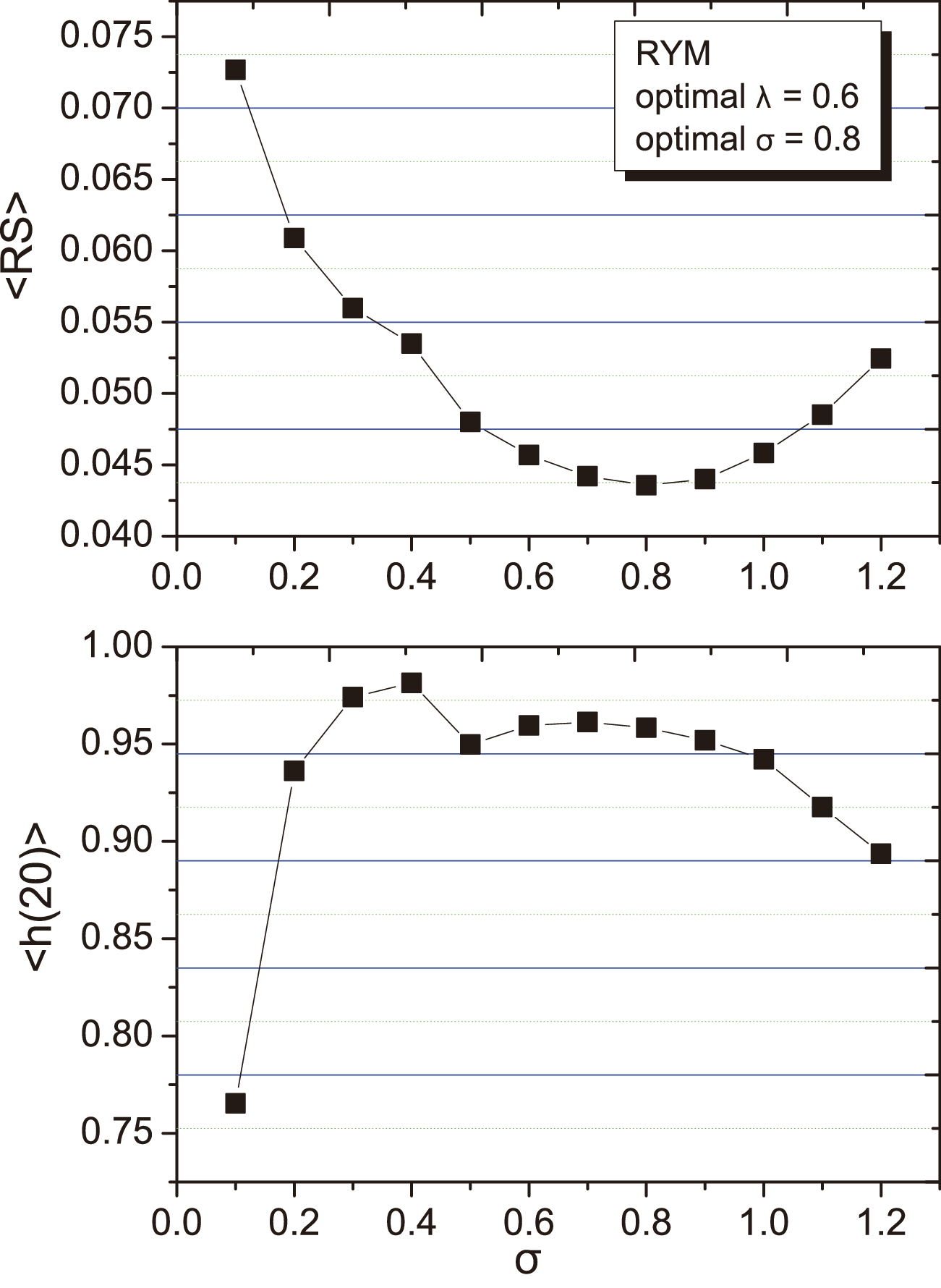}

\caption{The changes of ranking score and hamming distance when
traversing parameter $\sigma$, where $\lambda$ is tuned to be
optimal for every $\sigma$ value.}\label{fig:3}
\end{figure*}

By replacing RA with ERA index in the transfer equation of HHP
model, we get the enhanced HHP (Eh-HHP for short) model.
Fig.\ref{fig:3} plots the values of two typical measures, ranking
score and hamming distance, against parameter $\sigma$ traversed
from 0.1 to 1.2, where the other parameter $\lambda$ is tuned to be
the optimal value for the corresponding $\sigma$. We can see that in
a large range of $\sigma < 1$, the ranking score (hamming distance)
is smaller (larger) than that of $\sigma = 1$ for three data sets,
which means that the accuracy and diversity of \textcolor{red}{Eh-HHP} model will be
greatly improved than original HHP if $\sigma$ is tuned delicately.

Specifically, the ranking score reaches its optimal value when
$\sigma = 0.7$, $\sigma = 0.6$, and $\sigma = 0.8$ for MovieLens,
Netflix and RYM, respectively. Table.3 summarizes the detailed
statistics of recommending results for HHP and Eh-HHP models on
three data sets. We can see all the three metrics of Eh-HHP are
improved compared with HHP. Specially, the ranking score and hamming
distance are increased by $15.23\%$ and $17.3\%$ on Netflix, while
the performance improvements on MovieLens and RYM are not that
remarkable, which is in line with our \textcolor{red}{conjecture} in the above
subsection.

\begin{table*}[!hbp]
\label{tab:4}
\begin{center}
\caption{The recommendation performance of HHP and Eh-HHP models on
three data sets. }
\begin{tabular}{lcccccr}
\hline
Data & Methods & $\sigma_{opt}$ & $\lambda_{opt}$ &$RS$ & $ep(20)$ & $h(20)$\\
\hline
\multirow{2}*{MovieLens}
& HHP & &0.86&0.0923&26.33&0.9027\\
& Eh-HHP &0.7&0.58&\textbf{0.0894}&\textbf{27.95}&\textbf{0.9187}\\
\hline \multirow{2}*{Netflix}
& HHP& &0.83&0.04471&84.89&0.7559\\
& Eh-HHP &0.6&0.51&\textbf{0.0379}&\textbf{90.09}&\textbf{0.8866}\\
\hline \multirow{2}*{RYM}
& HHP& &0.76&0.04557&119.2&0.9369\\
& Eh-HHP &0.8&0.59&\textbf{0.0435}&\textbf{123.70}&\textbf{0.9552}\\
\hline
\end{tabular}
\end{center}
\end{table*}

\begin{table*}[!hbp]
\label{tab:result}
\begin{center}
\caption{The percentage improvement of metrics by introducing ERA in
ProbS, HeatS and HHP on three data sets. }
\begin{tabular}{l c c c c c r}
\hline
Methods& DataSets  & $\sigma_{opt}$ & $RS$ & $ep(20)$ & $h(20)$ \\
\hline \multirow{3}*{Eh-ProbS}
& MovieLens &0.4&3.24\%&3.3\%&4.9\%\\
& Netflix &0.5&3.32\%&0.7\%&12.45\%\\
& RYM &0.8&0.61\%&-1.6\%&0.61\%\\
\hline \multirow{3}*{Eh-HeatS}
& MovieLens &1.1&30.1\%&331.1\%&3.43\%\\
& Netflix &1.2&51.2\%&$\gg$100\%&-10.3\%\\
& RYM &1.1&17.6\%&26.4\%&0\%&\\
\hline \multirow{3}*{Eh-HHP}
& MovieLens &0.7&3.14\%&6.15\%&1.8\%\\
& Netflix &0.6&15.23\%&6.13\%&17.3\%\\
& RYM &0.8&4.5\%&3.78\%&1.95\%\\
\hline
\end{tabular}
\end{center}
\end{table*}

We further investigate the effect of ERA on ProbS and HeatS. Table.4
shows the percentage improvement of three metrics by enhancing the
RA similarities in three diffusion-like models ProbS, HeatS, and
HHP. It can be clearly observed that the enhancement of RA
similarity performs different on three models. For ProbS, which is
widely known as an accuracy-favored method, the enhanced similarity
brings an improvement on all the metrics on $MovieLens$ and
$Netflix$ data sets, but has unnoticed effect on $RYM$. For HeatS,
which is famous for its extremely high diversity and terrible
accuracy, the improvements of accuracy metrics on three data sets
are remarkable. However, the diversity of Eh-HeatS is decreased
compared with HeatS, but still better than Eh-ProbS. The
optimal $\sigma$ values of Eh-ProbS and Eh-HHP are all between 0 and
1, while for HeatS it lies on $\sigma>1$, which means that the
optimal $\sigma$ values depend not only on the data set but also on
the original model.

\begin{table*}[!hbp]
\label{tab:final result}
\begin{center}
\caption{Performance comparison of diffusion-like models on Netflix
data set.}
\begin{tabular}{lccccr}
\hline
  Methods & $\sigma_{opt}$ & $\lambda_{opt}$ &$RS$ & $ep(20)$ & $h(20)$\\
  \hline
 ProbS & n/a & n/a &0.050&74.524&0.5535\\
 Eh-ProbS&0.5 &n/a &0.048&74.017&0.6224\\
 \hline
 HeatS & n/a & n/a &0.106&0.098&0.8324\\
 Eh-HeatS&1.2&n/a&0.052&75.975&0.7464\\
 \hline
 HHP& n/a &0.83&0.045&84.895&0.7559\\
 Eh-HHP &0.6&0.51&\textbf{0.0379}&\textbf{90.090}&\textbf{0.8866}\\
 \hline
 BHC & n/a &0.85 &0.048&81.029&0.7602\\
 Eh-BHC&0.6&0.55&0.041&87.776&0.8469\\
 \hline
 BD & n/a & 0.77 & 0.039 & 82.537 & 0.8502\\
 Eh-BD&0.7&0.58&\textbf{0.0378}&88.344&0.8793\\
\hline
\end{tabular}
\end{center}
\end{table*}

A summary of performance comparison of several diffusion-like models
with the enhanced ones on Netflix data set are illustrated in
Table.5. The optimal parameters $\sigma$ and $\lambda$ are subject to the lowest ranking score, and the other two metrics $ep(20)$ and $h(20)$ are
calculated accordingly. We can see that almost all the enhanced models perform better than the original ones on both accuracy and diversity except Eh-HeatS which we have explained above. Clearly, Eh-HHP outperforms other enhanced algorithms over all three metrics except the case of Eh-BD in $RS$ where they are statistically comparable. We have known that without enhanced similarity, BD is the best among the five algorithms, while with ERA similarity, Eh-HHP is better than Eh-BD, i.e., the effect of ERA similarity on HHP is greater than that on BD incorporating with the effect of $\lambda$. We can further investigate that why the performance of Eh-BD is not remarkable on either accuracy or diversity compared with BD later on.

Most of the existing variants of ProbS and
HeatS try to diffuse more resources to unpopular and niche objects
to match the personalized interests of users, or to assign more
resources to popular objects to satisfy the tastes of the majority
of users. However, our idea changes essentially the distribution of
the similarities of objects, which can be applied into most of the
existing diffusion-like recommendation models without increasing the
computation complexity.

\section{Conclusions and Discussions}
An important challenge of recommender systems is how to accurately
recommend the unpopular objects in the absence of enough profile
information, without degradation of recommendation accuracy of
popular objects. In this paper, we proposed to enhance the RA
similarity in diffusion-like recommendation models to improve
simultaneously both of accuracy and diversity. Experimental results
demonstrated the effectiveness of this modification on ProbS, HeatS
and HHP on three typical data sets. Different from the power-law
distribution of original RA similarity, the distribution of ERA
similarity is more even, thus the similarity of a large proportion
pairs of unpopular objects are  increased, while the similarity of a
few popular objects remains almost unchanged. That is why the
proposed method greatly reinforces the existed diffusion-like
models.

Although our method can improve the recommendation results, we still
lack of a full understanding of the effect from the point of view of
network topology, e.g., why the improvement of accuracy and
diversity is more remarkable on Netflix than that on MovieLens, and
why the optimal $\sigma$ for Eh-HeatS is larger than 1. Since the
transfer equation $w_{\alpha\beta}$ consists of other components
besides RA (ERA) similarity, the influences of these factors should
not be neglected. Generally speaking, our work can be analogously
applied to any similarity-based recommendation models. We hope that
this idea can shed a light on the exploration of relation between
the characteristics of network topology and recommendation results.

\section{Acknowledgments}
\label{ACknowledgments} The authors would like to express their
gratitude to Tao Zhou, Yan-Li Li and Wen-Jun Li for helpful
discussions and irradiative ideas. This work is supported by the
Natural Science Foundation of China (\#61300018), and Special
Project of Sichuan Youth Science and Technology Innovation Research
Team (\#2013TD0006).

\end{document}